\documentclass[a4paper,fleqn,usenatbib]{mnras}

\usepackage{newtxtext,newtxmath}
\usepackage[T1]{fontenc}
\usepackage{ae,aecompl}

\usepackage{graphicx}	% Including figure files
\usepackage{amsmath}	% Advanced maths commands
\usepackage{amssymb}	% Extra maths symbols

\usepackage{color}
\usepackage{fancybox}
\usepackage{subfigure}
\usepackage{rotating}
\usepackage{import}
\usepackage{pdfpages}
\usepackage{xcolor}
\usepackage{hyperref}
\usepackage{verbatim} 
\usepackage{CJK}

\newcommand{\MgII}{Mg\,{\small II}}

\newcommand{\hbeta}{H{$\beta$}}
\newcommand{\halpha}{H{$\alpha$}}

\newcommand{\OII}{[O\,{\small II}]}
\newcommand{\OIIab}{[O\,{\small III}]\,$\lambda\lambda$3727,3729}
\newcommand{\OIII}{[O\,{\small III}]}
\newcommand{\OIIIa}{[O\,{\small III}]\,$\lambda$4959}
\newcommand{\OIIIb}{[O\,{\small III}]\,$\lambda$5007}
\newcommand{\OIIIab}{[O\,{\small III}]\,$\lambda\lambda$4959,5007}
\newcommand{\NII}{[N\,{\small II}]}

\newcommand{\objfull}{DES~J021822.52$-$043035.88}
\newcommand{\obj}{DES~J0218$-$0430}

\usepackage[switch,columnwise]{lineno}
\title[DES z=0.823 Low-Mass AGN]{Dark Energy Survey Identification of A Low-Mass Active Galactic Nucleus at Redshift 0.823 from Optical Variability}

\author[Guo et al.]{
Hengxiao~Guo,$^{1,2}$ %\thanks{E-mail: hengxiao@illinois.edu;colinjb2@illinois.edu;xinliuxl@illinois.edu}
Colin~J.~Burke,$^{1,2}$
Xin~Liu,$^{1,2}$
Kedar~A.~Phadke,$^{1}$
Kaiwen~Zhang,$^{3}$
\newauthor
Yu-Ching~Chen,$^{1,2}$
Robert~A.~Gruendl,$^{1,2}$
Christopher~Lidman,$^{4}$
Yue~Shen,$^{1,2}$%\thanks{Alfred P. Sloan Research Fellow}
\newauthor
Eric~Morganson,$^{2}$
Michel~Aguena,$^{5,6}$
Sahar~Allam,$^{7}$
Santiago~Avila,$^{8}$
Emmanuel~Bertin,$^{9,10}$
\newauthor
David~Brooks,$^{11}$
Aurelio~Carnero~Rosell,$^{12}$
Daniela~Carollo,$^{13}$
Matias~Carrasco~Kind,$^{1,2}$
\newauthor
Matteo~Costanzi,$^{14,15}$
Luiz~N.~da Costa,$^{6,16}$
Juan~De~Vicente,$^{12}$
Shantanu~Desai,$^{17}$
\newauthor
Peter~Doel,$^{11}$
Tim~F.~Eifler,$^{18,19}$
Spencer~Everett,$^{20}$
Juan~Garc\'ia-Bellido,$^{8}$
\newauthor
Enrique~Gaztanaga,$^{21,22}$
David~W.~Gerdes,$^{23,24}$
Daniel~Gruen,$^{25,26,27}$
Julia~Gschwend,$^{6,16}$
\newauthor
Gaston~Gutierrez,$^{7}$
Samuel~R.~Hinton,$^{28}$
Devon~L.~Hollowood,$^{20}$
Klaus~Honscheid,$^{29,30}$
\newauthor
David~J.~James,$^{31}$
Kyler~Kuehn,$^{32,33}$
Marcos~Lima,$^{5,6}$
Marcio~A.~G.~Maia,$^{6,16}$
\newauthor
Felipe~Menanteau,$^{1,2}$
Ramon~Miquel,$^{34,35}$
Anais~M\"oller,$^{36}$
Ricardo~L.~C.~Ogando,$^{6,16}$
\newauthor
Antonella~Palmese,$^{7,37}$
Francisco~Paz-Chinch\'{o}n,$^{38,2}$
Andr\'{e}s~A.~Plazas,$^{39}$
Anita~K.~Romer,$^{40}$
\newauthor
Aaron~Roodman,$^{26,27}$
Eusebio~Sanchez,$^{12}$
Vic~Scarpine,$^{7}$
Michael~Schubnell,$^{24}$
\newauthor
Santiago~Serrano,$^{21,22}$
Mathew~Smith,$^{41}$
Marcelle~Soares-Santos,$^{42}$
Natalia~E.~Sommer,$^{4}$
\newauthor
Eric~Suchyta,$^{43}$
Molly~E.~C.~Swanson,$^{2}$
Gregory~Tarle,$^{24}$
Brad~E.~Tucker,$^{4}$
\newauthor
Tamas~N.~Varga$^{44,45}$
(DES Collaboration)
}

\usepackage{eso-pic}% http://ctan.org/pkg/eso-pic

\begin{document}
%\begin{CJK*}{UTF8}{gbsn}
\label{firstpage}
\pagerange{\pageref{firstpage}--\pageref{lastpage}}
\maketitle

%\end{CJK*}

\begin{abstract}
We report the identification of a low-mass AGN, \obj , in a redshift $z = 0.823$ galaxy in the Dark Energy Survey (DES) Supernova field. We select \obj\ as an AGN candidate by characterizing its long-term optical variability alone based on DES optical broad-band light curves spanning over 6 years. An archival optical spectrum from the fourth phase of the Sloan Digital Sky Survey shows both broad \MgII\ and broad \hbeta\ lines, confirming its nature as a broad-line AGN. Archival XMM-Newton X-ray observations suggest an intrinsic hard X-ray luminosity of $L_{{\rm 2-12\,keV}}\approx 7.6\pm0.4\times10^{43}$ erg s$^{-1}$, which exceeds those of the most X-ray luminous starburst galaxies, in support of an AGN driving the optical variability. Based on the broad \hbeta\ from SDSS spectrum, we estimate a virial BH mass of $M_{\bullet}\approx10^{6.43}$--$10^{6.72}M_{\odot}$ (with the error denoting the systematic uncertainty from different calibrations), consistent with the estimation from OzDES, making it the lowest mass AGN with redshift $>$ 0.4 detected in optical. We estimate the host galaxy stellar mass to be $M_{\ast}\approx10^{10.5\pm0.3}M_{\odot}$ based on modeling the multi-wavelength spectral energy distribution. \obj\ extends the $M_{\bullet}$--$M_{\ast}$ relation observed in luminous AGNs at $z\sim1$ to masses lower than being probed by previous work. Our work demonstrates the feasibility of using optical variability to identify low-mass AGNs at higher redshift in deeper synoptic surveys with direct implications for the upcoming Legacy Survey of Space and Time at Vera C. Rubin Observatory.
\end{abstract}

%It was targeted as a quasar candidate by the SDSS-IV/eBOSS survey based on its optical/MIR color but was not included in the SDSS DR14 quasar catalog due to its low luminosity ($M_i{=}-20.5$)
%Archival XMM-Newton X-ray observations suggest a hard X-ray luminosity of $L_{{\rm 2-10\,keV}}\approx10^{43}$ erg s$^{-1}$ (around peak) and a high level of X-ray variability, in support of an AGN as the origin for the optical variability.
% Are we sure that the X-ray variability is significant?


%%%%%%%%%%%%%%%%%%%%%%%%%%%%%%%%%%%%%%%%%%%%%%%%%%

%%%%%%%%%%%%%%%%% BODY OF PAPER %%%%%%%%%%%%%%%%%%
%%%%%%%%%%%%%%%%%%%%%%%%%%%%%%
\section{Introduction}\label{sec:intro}

Supermassive BHs (SMBHs) as massive as $\sim$1--10 billion solar masses were already formed when the universe was only a few hundred Myr old \citep[e.g.,][]{Fan2001,Wu2015,Banados2018}. How they were able to form so quickly is an outstanding question in cosmology \citep{Volonteri2010}. At least three channels have been proposed for the formation of the seeds of SMBHs: pop III stellar remnants \citep[e.g.,][]{Madau2001}, direct collapse \citep[e.g.,][]{Haehnelt1993,Bromm2003,Begelman2006}, or star cluster evolution \citep[e.g.,][]{Gurkan2004,PortegiesZwart2004}. Finding small BH seeds directly in the early universe represents a major goal of future facilities \citep[e.g.,][]{TheLynxTeam2018}. 

The occupation fraction of BHs in local dwarf galaxies \citep[i.e., $M_{\ast} < 10^{10} M_{\odot},$][]{Greene2019} and their mass functions hold the fossil record for understanding the mechanisms of seed formation \citep[e.g.,][]{Greene2012,Reines2016}. There is growing evidence for the existence of intermediate mass BHs \citep[IMBHs, $M_{\bullet} = 10^2 - 10^6M_{\odot},$][]{Greene2019}, including in some globular cluster centers, ultra-luminous X-ray sources (ULXs), and the center of dwarf galaxies \citep[e.g.,][]{Mezcua2017}. However, most of the existing evidence is limited to the low-redshift ($z<0.15$) universe. Recently, \citet{Mezcua2019NA} pointed out a problem of using local dwarf galaxies as the hosts for BH seeds, which may be contaminated by mergers and/or AGN feedback and therefore may not be the ideal fossil record for studying seed formation. This underscores the importance of finding small BHs at higher redshift, because they are more ``pristine'' (i.e., have gone through fewer mergers and feedback) than those at lower redshift.

\begin{figure*}
 \centering
 \includegraphics[width=1\textwidth]{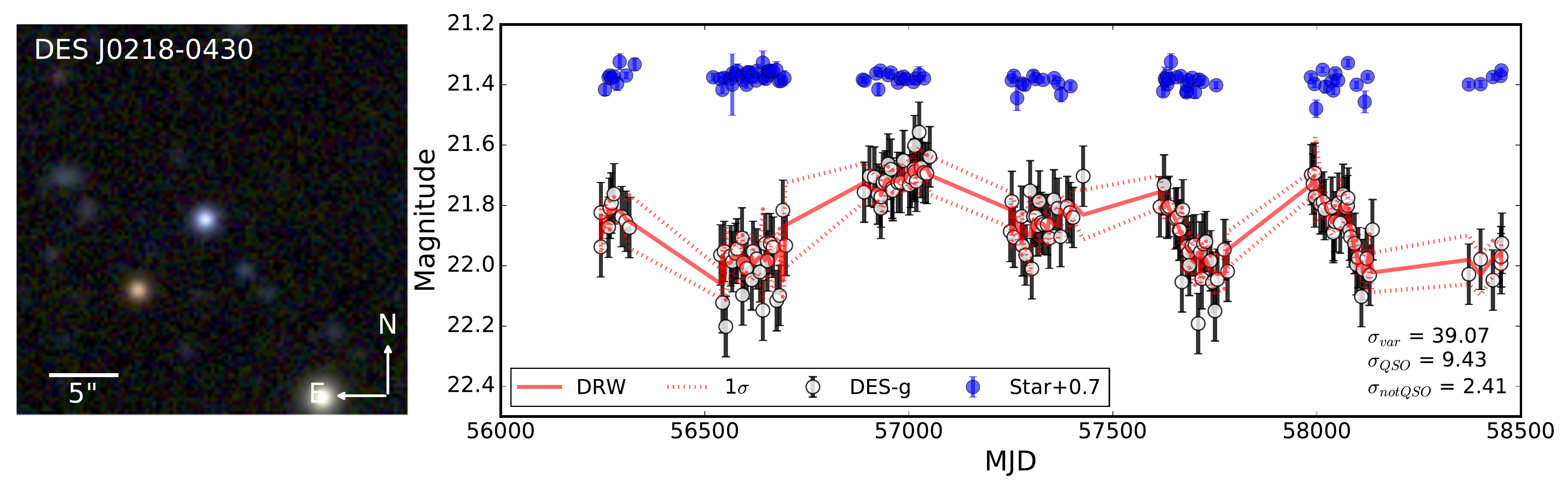}
    \caption{Left panel: DES $gri$-color composite image (with a 30$''\times$30$''$ field of view) for \obj . 
    %The red circle is 1\farcs3 in radius, representing the extraction aperture for the light curves. 
   Right panel: DES $g$-band PSF magnitude light curves for \obj\ (open filled circles) and a field star (blue filled circles) for comparison. The best-fit model for \obj\ (the red solid) and the 1 $\sigma$ confidence levels (the red dashed) assume a damped random walk \citep{Kelly2009}. Labeled in the lower right are the variability significance, QSO significance and non-AGN variability significance. See \S \ref{subsec:variability} for details. 
    }\label{fig:lc}
\end{figure*}

Previously, the best strategy for identifying low-mass AGNs at higher redshift was using deep X-ray surveys, such as the Chandra deep fields (CDF) \citep[e.g.][]{Fiore2012,Young2012,Luo2017,Xue2017} and the COSMOS survey \citep{Civano2012} (also see our Figure \ref{fig:redshift_mass}). For example, \citet{Luo2017} detected $\sim$1000 objects in CDF-South (484.2 arcmin$^2$) with total 7 Ms exposure time. 711 are AGNs based on the X-ray and multiwavelength properties. However, deep X-ray surveys are expensive and often plagued by contamination from star formation and/or X-ray binaries. Radio searches for low-mass AGNs in nearby dwarf galaxies have also been conducted with NSF's Karl G. Jansky Very Large Array high resolution observations \citep[e.g.,][]{Reines2020}, although they are subject to the low detection rate of radio cores of AGNs. Alternatively, optical color selection is much less expensive but is biased against smaller BHs and/or lower Eddington ratios. Optical emission line selection may miss AGNs with line ratios dominated by star formation \citep[e.g.,][]{Baldassare2016,Agostino2019}, particularly in low-mass galaxies without sufficient spectral resolution \citep{Trump2015}. Furthermore, the standard optical narrow emission line diagnostics used to identify SMBHs may fail when the BH mass falls below ${\sim}10^4M_{\odot}$ for highly accreting IMBHs and for radiatively inefficient IMBHs with active star formation, because the enhanced high-energy emission from IMBHs could result in a more extended partially ionized zone compared with models for SMBHs, producing a net decrease in the predicted \OIII /\hbeta\ and \NII /\halpha\ emission line ratios \citep[e.g.,][]{Cann2019}.

In this work, we present the identification of \objfull\ (hereafter \obj\ for short) as a low-mass AGN at $z{=}0.823$ by characterizing its optical variability based on sensitive, long-term light curves from the Dark Energy Survey \citep[DES;][]{Flaugher2005,TheDarkEnergySurveyCollaboration2005,DarkEnergySurveyCollaboration2016} Supernova (SN) fields \citep{Kessler2015}. It serves as a proof of principle for identifying low-mass AGNs (i.e., $M_{\bullet} \lesssim 10^{6} M_{\sun} $) at intermediate and high redshift using deep synoptic surveys with important implications for the Rubin Observatory Legacy Survey of Space and Time \citep[LSST;][]{Ivezic2019}.

Compared to other methods, variability searches should be more sensitive to AGNs with lower Eddington ratios given the anti-correlation between Eddington ratio and optical variability \citep{MacLeod2010,Ai2010,Guo2014,Rumbaugh2017,Lu2019,Sanchez-Saez2018}. Recently, \citet{Baldassare2018} selected several low-mass AGN candidates in the Sloan Digital Sky Survey \citep{York2000} Stripe 82 \citep[SDSS-S82;][]{Ivezic2007,SDSSDR7}, but the sample is limited to $z<0.15$ by the sensitivity of SDSS-S82 light curves. Compared to SDSS-S82, DES-SN provides a factor of 10 increase in single-epoch imaging sensitivity. The higher sensitivity is crucial for discovering AGNs with lower masses at higher redshift.

%Currently the only potential rival for us is PanSTARRS but DES is still ~2 mag deeper (4 m vs 1.8 m telescope). HSC used to have a supernova component but unfortunately didn't make it due to weather. ZTF is getting lots of the low-hanging fruits for LSST, but it just can't do anything so faint!

%It is a big deal to have a direct virial mass measurement for a small BH at intermediate redshift! Identifying small BHs and measuring their masses are two different things, and the latter is even more challenging. most of the estimates come from either BH-host scaling relation or from X-ray luminosity assuming some fiducial Eddington ratio. The latter is wildly uncertain (a guess at best); with the former one will not be able to do any study on the BH host relation. Because \obj\ actually has a mass measurement from broad lines (even though the virial method itself has caveats), we would be able to study its BH-host scaling relation, if we can actually resolve and measure its host galaxy.

Our main new findings include:

\begin{enumerate}
\item Identification of a low-mass AGN based on optical variability alone. This represents the first low-mass AGN identified from optical variability at intermediate redshift.

\item Confirmation that the optical variability is driven by an AGN based on optical spectroscopy, high hard X-ray luminosity, and broad band spectral energy distribution (SED).

\item Estimation of the BH mass $M_{\bullet}$ using the virial method. Combined with the stellar mass estimate $M_{\ast}$ from SED modeling, this puts \obj\ on the $M_{\bullet}$--$M_{\ast}$ relation in AGN at intermediate redshift and extends it to lower masses than probed by previous work.  

\item Demonstration that variability searches based on sensitive, long-term optical light curves from deeper synoptic surveys can indeed identify low-mass AGNs at higher redshift \citep[see also][for a recent study based on NIR variability]{Elmer2020}.

\end{enumerate}

The paper is organized as follows. Section \ref{sec:data} describes the observations and data analysis that identify \obj\ as a candidate low-mass AGN from optical variability and provides confirmation of its AGN nature based on optical spectroscopy and multi-wavelength properties. Section \ref{sec:result} presents our results on the estimation of its virial BH mass and the host galaxy stellar mass. We discuss the implications of our results in Section \ref{sec:discuss} and conclude in \ref{sec:sum}. A concordance $\Lambda$CDM cosmology with $\Omega_m = 0.3$, $\Omega_{\Lambda} = 0.7$, and $H_{0}=70$ km s$^{-1}$ Mpc$^{-1}$ is assumed throughout. We use the AB magnitude system \citep{Oke1974} unless otherwise noted.

\section{Observations and Data Analysis}\label{sec:data}

\subsection{Variability Characterization}\label{subsec:variability}

To distinguish AGN variability from variable stellar sources (e.g., stars, SNe), we follow the method of \citet{Butler2011}, which represents an easy to implement method for selection of quasars using single-band light curves. We focus on the $g$ band given that AGNs tend to show larger variability amplitudes in bluer bands \citep{Ulrich1997}. The \citet{Butler2011} method first uses the damped random walk model \citep{Kelly2009} to parameterize the ensemble quasar structure function in SDSS-S82. Then, based on this empirical variable QSO structure function, they classify individual light curves into variable/ non-variable objects and QSO/non-QSO with no parameter fitting. 

%The \citet{Butler2011} method uses the damped random walk model of \citet{Kelly2009} and has been demonstrated with spectroscopically confirmed quasars based on SDSS-S82 light curves. 

The variability classification is based on two statistics, one describing the fit confidence and the other describing the false alarm probability (FAP), which is tuned to achieve high quasar detection fractions given an acceptable FAP. More specifically, we use the software qso\_fit\footnote{\url{http://butler.lab.asu.edu/qso\_selection/index.html}} to model the light curve and quantify if a source is variable and if yes, whether the variability is characteristic of AGN. We calculate the following statistics:
\begin{enumerate}
\item $\sigma_{{\rm var}}$: the significance that a source is variable,
\item $\sigma_{{\rm QSO}}$: the significance that a source is variable and that the fit to the damped random walk model is statistically preferred over that to a randomly variable, and
\item $\sigma_{{\rm notQSO}}$: the significance that a source is variable but is not characteristic of AGN. This parameter is usually anti-correlated with $\sigma_{{\rm QSO}}$, lending further support to the AGN classification.
\end{enumerate}

Although optimized for quasar variability, the \citet{Butler2011} method has been demonstrated to find variability in dwarf galaxies \citep{Baldassare2018}.

\subsection{Target Selection Using the Dark Energy Survey}\label{subsec:target}

%Describe DES and DES-SN.

DES (Jan 2013--Jan 2019) was a wide-area 5000 deg$^2$ survey of the Southern Hemisphere in the $grizY$ bands. It used the Dark Energy Camera \citep{Flaugher2015,Bernstein2017a} with a 2.2 degree diameter field of view mounted at the prime focus of the Victor M. Blanco 4m telescope on Cerro Tololo in Chile. The typical single-epoch 5$\sigma$ point source depths achieved with six-year's data are $g$=24.7, $r$=24.5, $i$=23.9, $z$=23.3, and $Y$=21.8 mag \citep[$\sim$0.4 mag deeper than three-year's data,][]{Abbott2018}, much deeper than other surveys of larger area (e.g., SDSS-S82 and PanSTARRS1). The data quality varies due to seeing and weather variations. DES absolute photometric calibration has been tied to the spectrophotometric Hubble CALSPEC standard star C26202 and has been placed on the AB system \citep{Oke1983}. The estimated single-epoch photometric statistical precision is 7.3, 6.1, 5.9, 7.3, 7.8 mmag in the $grizY$ bands \citep{Abbott2018}. DES contains a 30 deg$^2$ multi-epoch survey DES-SN to search for SNe Ia. It observed in eight ``shallow" and in two ``deep" fields, with the shallow and deep fields having typical nightly
point-source depths of 23.5 and 24.5 mag, respectively \citep{Kessler2015,Brout2019}. DES-SN has a mean cadence of $\sim$7 days in the $griz$ bands between mid-August through mid-February from 2013 to 2019.

We have selected \obj\ as a candidate low-mass AGN by characterizing its long-term optical variability based on DES Y6A1 data. Details of our sample selection will be presented in a forthcoming paper. We briefly describe the selection procedure as follows: 
\begin{enumerate}
\item We started from an internal DES variability catalog in the DES-SN fields. The catalog includes AGNs, SNe and artifacts. We applied the damped random walk model to the variable light curves to select AGN-like variability (see the details in \S \ref{subsec:variability}).
\item We have required that the stellar mass estimates are less than $10^{10}M_{\odot}$ based on mass-to-light ratios (M/L) inferred from broad-band colors \citep{Taylor2011} without more careful SED fitting (see below in \S \ref{subsec:host_mass}), assuming that low-mass AGNs usually reside in low-mass galaxies.
\end{enumerate}

This resulted in $\sim1,300$ ``low-mass'' AGN candidates, although the actual number of low-mass AGN candidates is likely to be much smaller considering that our simple color-derived M/L and stellar masses would have been significantly underestimated due to contamination from a blue AGN continuum. We then cross-matched the candidates with the Million Quasar Catalog\footnote{\url{https://heasarc.gsfc.nasa.gov/W3Browse/all/milliquas.html}}. \obj\ was the only object with both an X-ray detection and obvious broad emission lines with widths of $\sim$500--2000 $\rm km\ s^{-1}$ from the SDSS spectra. We have also found other low-mass AGN candidates which either show only narrow emission-line components in their SDSS spectra, or, with X-ray detections but have no available SDSS spectrum (and therefore without a virial mass estimate). Spectroscopic follow-up observations are still needed for those candidates to measure any broad emission-line components to confirm their AGN nature and to infer their virial black hole masses.

\begin{figure}
 \centering
 \includegraphics[width=0.5\textwidth]{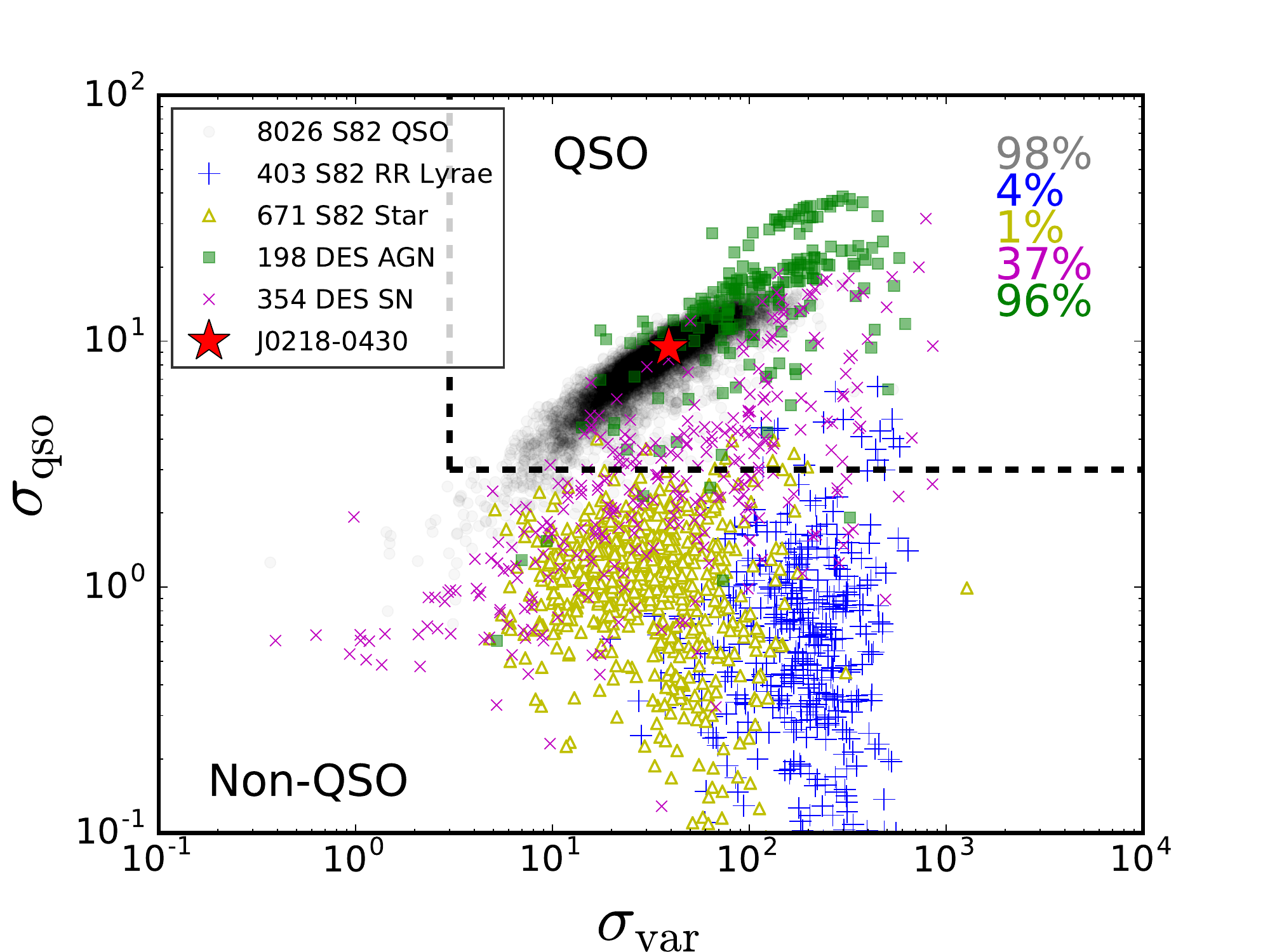}
    \caption{
QSO significance ($\sigma_{{\rm qso}}$) versus variability significance ($\sigma_{{\rm var}}$) for \obj . Also shown for context are spectroscopically confirmed quasars (grey dots) and stars (blue crosses and yellow triangles are for RR lyrae and non-variable stars, respectively) from the SDSS Stripe 82 and spectroscopically confirmed DES AGNs (green squares) and SNe (magenta crosses) from the OzDES survey \citep{Yuan2015}.
Numbers indicate the fraction of objects among each population classified as ``QSO'' using the criteria $\sigma_{{\rm var}}{>}3$ and $\sigma_{{\rm qso}}{>}3$ \citep{Butler2011}.
\obj\ is located in the region in which reside by most SDSS Stripe 82 quasars and DES AGN.
    }\label{fig:variability_selection}
\end{figure}

%Describe how \obj\ was selected as a candidate low-mass AGN.
Figure \ref{fig:lc} shows the $g$-band light curve of \obj\ (located in a shallow field) using the point-spread function (PSF) magnitudes. There are 142 epochs (175 sec/epoch) of observations in total. Unlike low-mass AGNs at lower redshift, the host galaxy of \obj\ is unresolved in DES. We therefore adopt the PSF magnitude photometry which is most appropriate for unresolved sources.

\begin{figure}
 \centering
 \includegraphics[width=0.5\textwidth]{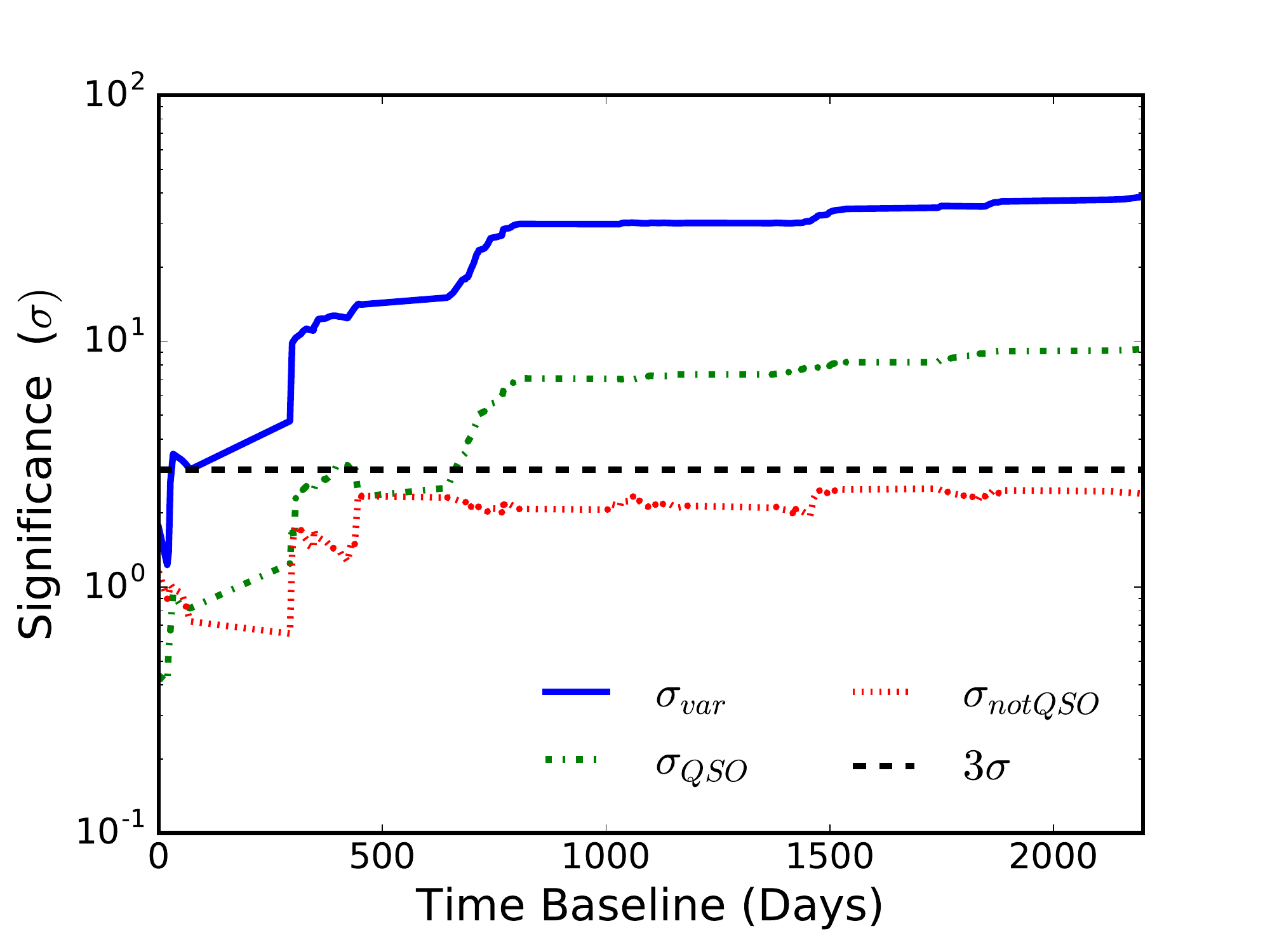}
    \caption{
The dependence of variability significance ($\sigma_{{\rm var}}$), QSO significance ($\sigma_{{\rm QSO}}$), and non-QSO significance ($\sigma_{{\rm notQSO}}$) on the total light curve baseline for \obj . 
    }\label{fig:baseline}
\end{figure}

Figure \ref{fig:variability_selection} shows $\sigma_{{\rm QSO}}$ versus $\sigma_{{\rm var}}$ for \obj\ compared against spectroscopically confirmed SDSS quasars and stars as well as DES AGN and SNe spectroscopically confirmed by OzDES. It demonstrates that \obj\ is classified as an AGN based on its characteristic optical variability at a high significance (with $\sigma_{{\rm var}}{\sim}$39 and $\sigma_{{\rm QSO}}{\sim}$9). It occupies the same subregion of parameter space as those of spectroscopically confirmed SDSS quasars and DES AGN. 

Figure \ref{fig:baseline} shows the dependence of variability significance ($\sigma_{{\rm var}}$), QSO significance ($\sigma_{{\rm QSO}}$), and non-QSO significance ($\sigma_{{\rm notQSO}}$) on the total light curve baseline $T$. While \obj\ can be classified as an AGN when $T{\gtrsim}$2 years, both $\sigma_{{\rm var}}$ and $\sigma_{{\rm QSO}}$ continue to increase with increasing $T$ until they start to saturate around $T{\sim}4$ years. This demonstrates the importance of a moderately long time baseline for AGN identification from optical variability.

\subsection{Optical Spectroscopy}\label{subsec:spec} 

\begin{figure*}
 \centering
 \includegraphics[width=0.8\textwidth]{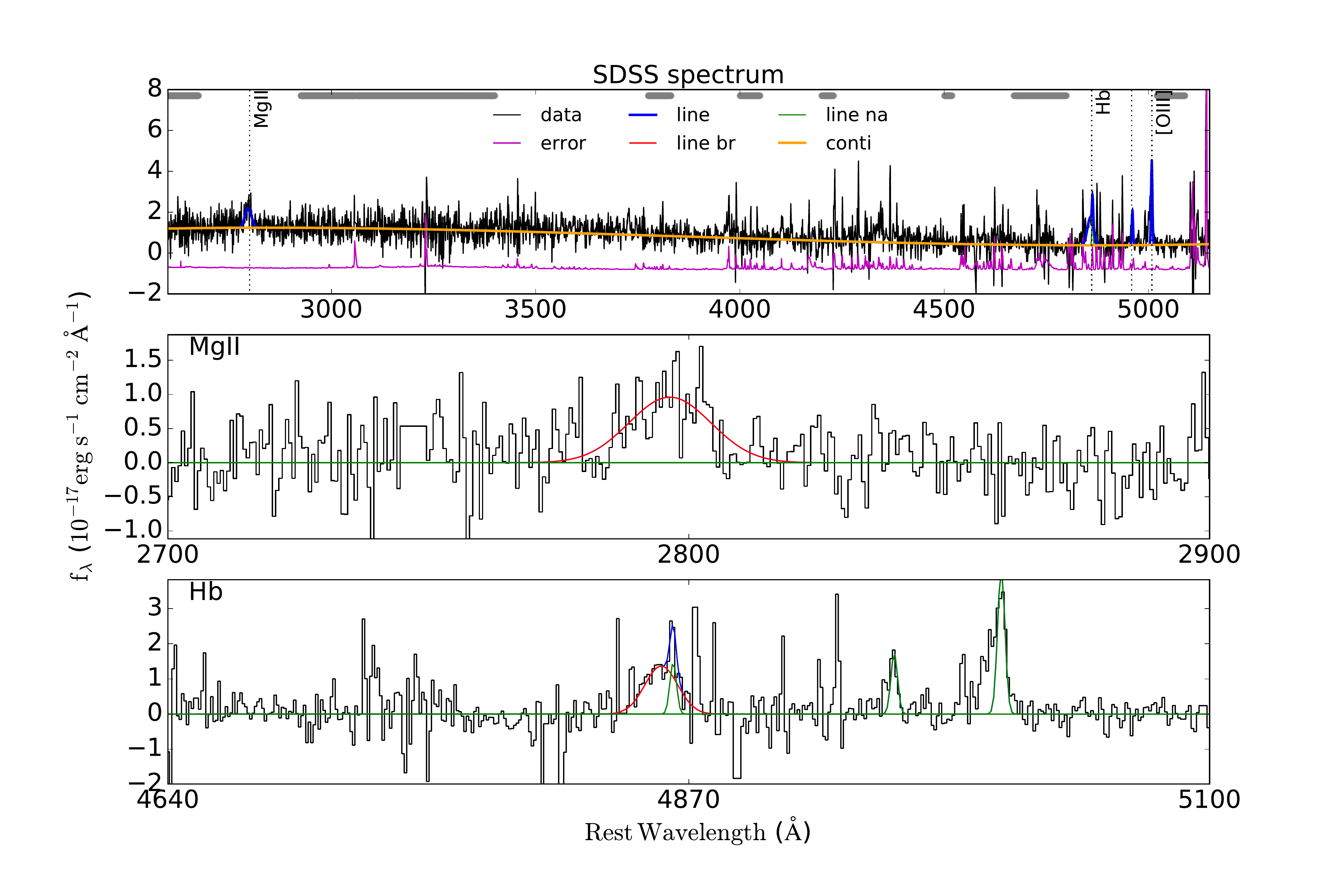}
 \includegraphics[width=0.8\textwidth]{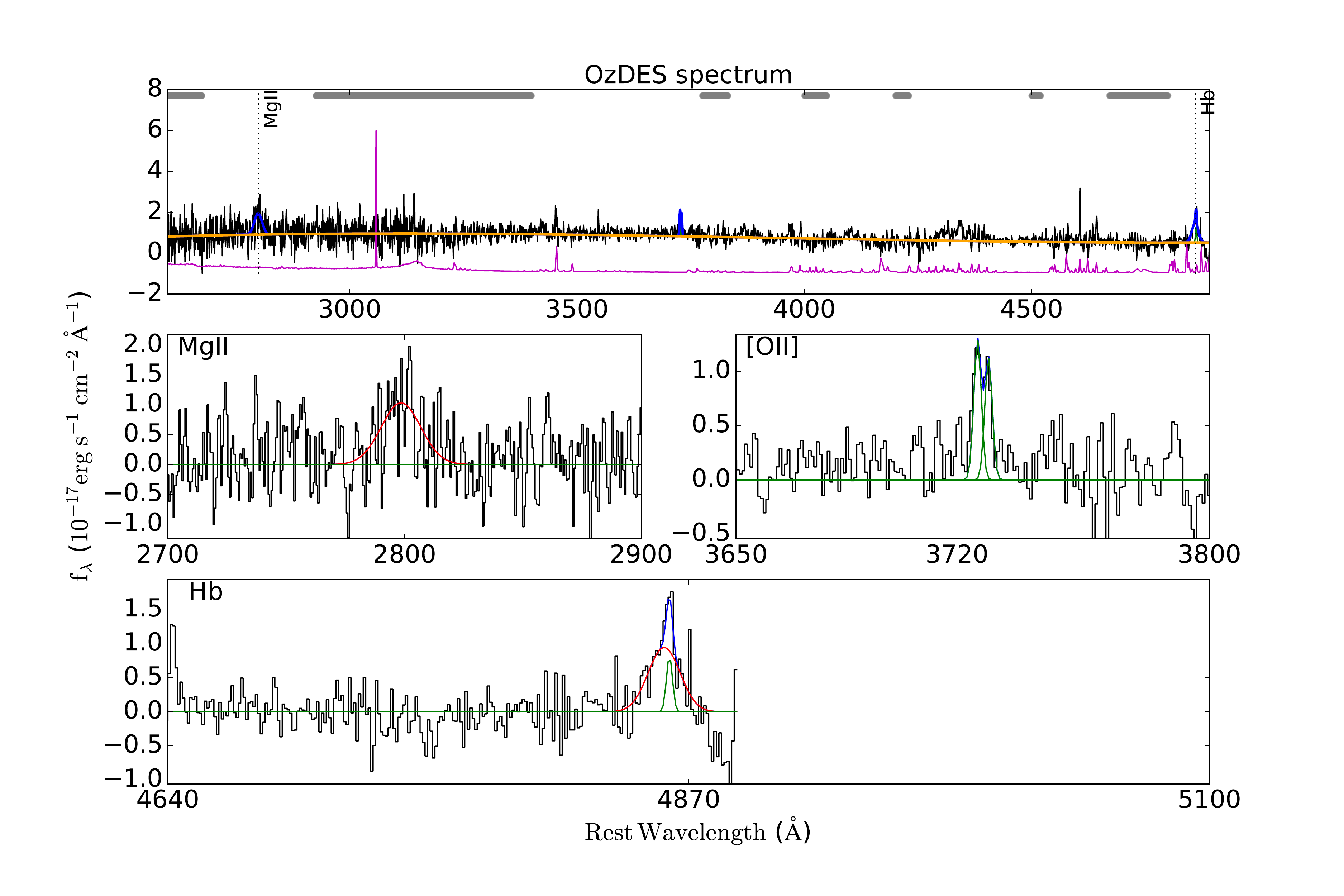}
    \caption{Optical spectrum for \obj\ from the SDSS-IV/eBOSS \& OzDES survey and our spectral modeling analysis. A global fitting is applied to the spectrum having subtracted the host component in the upper panel. Power-law + 3-order polynomial and Gaussians are used to fit the continuum and emission lines, respectively. The grey bands on the top are line-free windows selected to determine the continuum emission. The error spectrum has been shifted vertically by $-1\times10^{-17}$ erg s$^{-1}$ cm$^{-2}$ {\AA}$^{-1}$ for clarity. The lower panels show the zoomed-in emission line regions of \MgII , \OII\ and \hbeta . Broad \MgII\ and broad \hbeta\ are both detected at the 2.1(2.0)$\sigma$ and 3.4(3.6)$\sigma$ significance levels, yielding virial BH masses $\sim$10$^{6.43}$--$10^{6.72}M_{\odot}$ ($\sim$10$^{6.40}$--$10^{6.69}M_{\odot}$) using \hbeta\ from SDSS (OzDES).}
    \label{fig:boss_spec}
\end{figure*}

Figure \ref{fig:boss_spec} (upper panel) shows the archival optical spectrum (Plate ID = 8124, Fiber ID = 690, and MJD = 56954) of \obj\ from SDSS-IV \citep{Blanton2017}. It was targeted as a quasar candidate by the eBOSS survey \citep{Dawson2016} based on its optical/MIR color and was included in the SDSS DR14 quasar catalog \citep{Paris2018}. Its luminosity is $M_{i}=-20.5$ mag, which is below the SDSS DR7 quasar catalog luminosity criterion \citep[$M_i<-22$ mag;][]{Schneider2010}. It is not included in DES OzDES quasar catalog by \citet{Tie2017}, which has $M_i<-22$ mag. Both broad \hbeta\ and broad \MgII\ emission are covered in the spectrum.

\obj\ was observed by OzDES\footnote{Australian Dark Energy Survey} twice, once during 2014, and again in 2018. Since 2013, OzDES has used the 2dF positioner and AAOmega spectrograph on the Anglo-Australian Telescope to obtain redshifts for tens of thousands of sources within the 10 deep fields of the Dark Energy Survey \citep{Yuan2015,Childress2017}. The spectra from 2014 and 2018 are combined and shown in Figure \ref{fig:boss_spec} (lower panel). The total integration time for the combined spectrum was 3 hours. Further details on how the data were obtained and processed can be found in \citet{Yuan2015}, \citep{Childress2017}, and Lidman et al.~(in preparation).

To determine the significance of the broad emission lines and to measure their profiles for virial BH mass estimates, we fit spectral models following the procedures as described in detail in \citet{Shen2019} using the software PyQSOFit\footnote{\url{https://github.com/legolason/PyQSOFit}} \citep{Guo2018}. The model is a linear combination of a power-law continuum, a 3rd-order polynomial (to account for reddening), 
a pseudo continuum constructed from Fe\,II emission templates, and single or multiple Gaussians for the emission lines. Since uncertainties in the continuum model may induce subtle effects on measurements for weak emission lines, we first perform a global fit to the emission-line free region to better quantify the continuum. We then fit multiple Gaussian models to the continuum-subtracted spectrum around the broad emission line region locally.

More specifically, we model the \MgII\ line using a combination of up to two Gaussians for the broad component and one Gaussian for the narrow component.
We impose an upper limit of 1200 km s$^{-1}$ for the FWHM of the narrow lines. For the \hbeta\ line, we use up to three Gaussians for the broad \hbeta\ component and one Gaussian for the narrow \hbeta\ component. We use two Gaussians for the \OIIIa\ and \OIIIb\ narrow lines. Considering the low S/N of the spectrum, we only fit single Gaussians to the \OIIIab\ lines with the flux ratio of the doublet tied to be $f_{{\rm 5007}}/f_{{\rm 4959}}=3$. The line widths of \OIII\ and narrow \hbeta\ are tied together. Fitting each \OIII\ line with two Gaussians instead (with an additional component to account for a possible blue wing often seen in \OIII ) does not improve the fit significantly. The resulting broad-line \hbeta\ width is relatively insensitive to our model choice for \OIII . For OzDES spectrum without \OIII , we use \OIIab\ instead, which is fitted with two Gaussians to decompose the narrow component of \hbeta . We use 100 Monte Carlo simulations to estimate the uncertainty in the line measurements.

Figure \ref{fig:boss_spec} shows our best-fit spectral model for \obj . Table \ref{tab:measurement} lists the spectral measurements for \obj . Both broad \hbeta\ and broad \MgII\ are detected. This confirms \obj\ as a broad-line AGN.

\begin{table*}
\centering
\setlength{\tabcolsep}{1mm}
\begin{tabular}{ccccccccccccc}
  \hline
  \hline
 & F$_{{\rm Mg\,II}}$ & F$_{{\rm H}\beta}$ & FWHM$_{{\rm Mg\,II}}$ & FWHM$_{{\rm H}\beta}$ & log L$_{{\rm 3000}}$ & log L$_{{\rm 5100}}$ & M$^{{\rm Mg\,II},\,{\rm M16}}_{\bullet}$ & M$^{{\rm Mg\,II},\,{\rm S11}}_{\bullet}$ &  M$^{{\rm Mg\,II},\,{\rm VO09}}_{\bullet}$ & M$^{{\rm H}\beta,\,{\rm M16}}_{\bullet}$ & M$^{{\rm H}\beta,\,{\rm VP06}}_{\bullet}$ &  M$^{{\rm H}\beta,\,{\rm MD04}}_{\bullet}$ \\
 Spectrum &\multicolumn{2}{c}{(10$^{-17}$ erg s$^{-1}$ cm$^{-2}$)} & (km s$^{-1}$) & (km s$^{-1}$) & (erg s$^{-1}$) & (erg s$^{-1}$) &  (log$M_{\odot}$) & (log$M_{\odot}$) & (log$M_{\odot}$) & (log$M_{\odot}$) & (log$M_{\odot}$) & (log$M_{\odot}$) \\
(1) & (2) & (3) & (4) & (5) & (6) & (7) & (8) & (9) & (10) & (11) & (12) & (13) \\
\hline
SDSS &18.7$\pm$3.3  & 23.5$\pm$1.5 & 1980$\pm$360 & 1060$\pm$130 & 43.69 & 43.52 & 7.36$\pm$0.12 & 7.14$\pm$0.14 & 7.30$\pm$0.15 & 6.63$\pm$0.14 & 6.72$\pm$0.11 & 6.43$\pm$0.11  \\
OzDES &21.6$\pm$4.1  & 16.5$\pm$0.9 & 2118$\pm$410 & 1025$\pm$80 &  43.77 & \dots & 7.47$\pm$0.24 & 7.25$\pm$0.15 & 7.40$\pm$0.15 & 6.60$\pm$0.12 & 6.69$\pm$0.07 & 6.40$\pm$0.07  \\
\hline
\end{tabular}
\caption{Spectral measurements and virial black hole mass estimates of \obj . Cols. 2 and 3: Broad emission line flux and 1$\sigma$ uncertainty from Monte Carlo simulations. Cols. 4 and 5: Full width at half maximum of the broad emission line and 1$\sigma$ uncertainty measured from our best-fit spectral model (\S \ref{subsec:spec} and Figure \ref{fig:boss_spec}). Cols. 6 and 7: Monochromatic continuum luminosities of the AGN component in our best-fit spectral model after subtracting the host galaxy contribution from SED modeling. Cols. 8--13: Virial BH mass estimates using the calibrations of \citet{Mejia-Restrepo2016} (M16), \citet{Shen2011} (S11), and \citet{Vestergaard2009} (VO09) for \MgII\ and those of \citet{Mejia-Restrepo2016}, \citet{Vestergaard2006} (VP06), and \citet{McLure2004} (MD04) for \hbeta\ (Equations \ref{eq:virialmass} and \ref{eq:virialcoeff}). We assume the same 5100\AA\ luminosity as that from the SDSS to calculate the BH mass from OzDES.
\label{tab:measurement}
}
\end{table*}

\subsection{Multi-wavelength Observations}\label{subsec:sed_data}

To estimate the host-galaxy stellar mass (see \S \ref{subsec:host_mass} below for details), we queried the archival SED data for \obj\ using the Vizier tool\footnote{\url{http://vizier.u-strasbg.fr/vizier/sed/}} within 3 arcsec following the procedures of \citet{Guo2020}. We adopt measurements from large systematic surveys to focus on a more homogeneous data set. These include the Galaxy Evolution Explorer \citep[GALEX;][]{Martin2005}, the Sloan Digital Sky Survey \citep[SDSS;][]{York2000}, the UKIRT Infrared Deep Sky Survey \citep[UKIDSS;][]{Lawrence2007}, the Wide-field Infrared Survey \citep[WISE;][]{Wright2010}, and the Spitzer Wide-Area Infrared Extragalactic survey \citep[SWIRE;][]{Rowan-Robinson2013}. When multi-epoch photometries are available, we take the mean value to quantify the average SED. We assume 20\% as the fiducial fractional uncertainty if a proper photometric error is not available. 

\obj\ is included in the Ninth Data Release of the fourth Serendipitous Source Catalog (4XMM-DR9) of the European Space Agency's (ESA) XMM-Newton observatory \citep{Rosen2016}. It was detected at $>$ 6$\sigma$ significance as a compact source in a 21 ks exposure on 2016 July 1. The EPIC 2--4.5 KeV and 4.5--12 KeV fluxes are $(2.10 \pm 1.54) \times 10^{-15}$ $\rm erg\ cm^{-2}\ s^{-1}$ and $(2.19 \pm 1.14)\times 10^{-14}$ $\rm erg\ cm^{-2}\ s^{-1}$ respectively, yielding $L_{\rm 2-12\,keV} = (7.6 \pm 0.4)\times 10^{43}$ $\rm erg\ s^{-1}$. The X-ray luminosity exceeds those of the most X-ray luminous starburst galaxies \citep[e.g.,][]{zezas01}, lending further evidence for its AGN nature driving the optical variability.

\begin{figure*}
 \centering
 \includegraphics[width=0.9\textwidth]{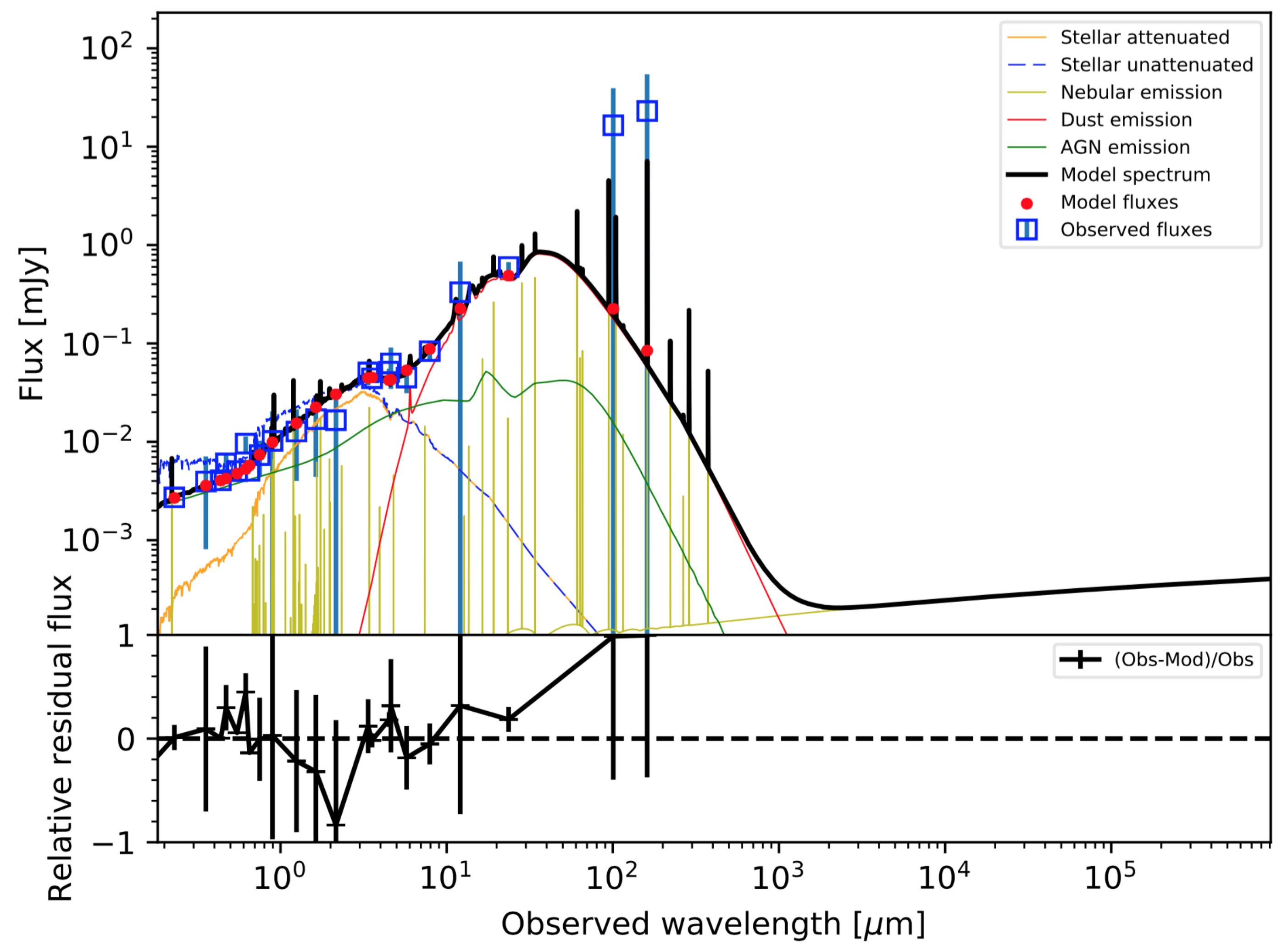}
    \caption{Spectral energy distribution modeling for \obj\ using CIGALE. All the photometry data from the Vizier service (see \S \ref{subsec:sed_data} for details) are shown as blue squares. The stellar unattenuated SED component is shown as the blue dotted line with the re-processed component shown as the solid orange line. Nebular emission is shown as the yellow solid line. The cold dust component is shown in red whereas the hot dust component from the AGN is shown in green. The best fit model is shown as the solid black line with residuals of observed and modeled flux values in the bottom panel. 
    }
    \label{fig:sed}
\end{figure*}

\section{Results}\label{sec:result}

%\subsection{Statistical Properties of the Optical Variability}\label{subsec:variability_stat}

%\subsection{X-ray Variability}\label{subsec:xray}
% i deleted this because it is unclear if there were any significant x-ray variability. we need to re-analyze xmm newton data.

\subsection{Black Hole Mass Estimation}\label{subsec:bh_mass}

We estimate the AGN BH mass using the single-epoch estimator assuming virialized motion in the broad-line region (BLR) clouds \citep{Shen2013}. With the continuum luminosity as a proxy for the BLR radius and the broad emission line width, characterized by the full width at half maximum (FHWM), as an indicator of the virial velocity, the virial mass estimate is given by

%\red{\begin{equation}\label{eq:virialmass}
%    \log \bigg(\frac{M_{\bullet}}{M_{\odot}} \bigg) = a + b \log \bigg(\frac{L_{\rm 5100}}{10^{44}\,{\rm erg\,s^{-1}}}\bigg) + 2 \log \bigg(\frac{\rm{FWHM}}{{\rm 10^{3}\, km\,s^{-1}}}\bigg),
%\end{equation}
%where $L_{\lambda}=L_{{\rm 3000}}$ for \MgII\ and $L_{\lambda}=L_{{\rm 5100}}$ for \hbeta\ derived from the global fitting. The calibration coefficients are:}
%\red{\begin{equation}\label{eq:virialcoeff}
%\begin{split}
%    (a,\,b) &= (6.955, \,0.599), \,  {\rm Mg\,II} \\
%    (a,\,b) &= (6.864, \,0.568), \, {\rm H}\beta .
%\end{split}
%\end{equation}}
%\red{We adopt the calibrations of \cite{Mejia-Restrepo2016} for \MgII\ and \hbeta . The coefficients $a$ and $b$ are initially empirically calibrated either against local reverberation mapped AGNs or internally among different lines, and then re-scaled based on the measurements of \hbeta\ and \MgII\ for about 30 AGNs at $\sim$1.5. Therefore, this BH estimator is more accurate for intermediate-redshift AGNs.}

\begin{equation}\label{eq:virialmass}
    \log \bigg(\frac{M_{\bullet}}{M_{\odot}} \bigg) = a + b \log \bigg(\frac{\lambda L_{\lambda}}{10^{44}\,{\rm erg\,s^{-1}}}\bigg) + 2 \log \bigg(\frac{\rm{FWHM}}{{\rm km\,s^{-1}}}\bigg),
\end{equation}

where $L_{\lambda}=L_{{\rm 3000}}$ for \MgII\ and $L_{\lambda}=L_{{\rm 5100}}$ for \hbeta . The coefficients $a$ and $b$ are empirically calibrated either against local reverberation mapped AGNs or internally among different lines. We adopt the calibrations of \citet{Mejia-Restrepo2016}\footnote{This calibration is based on a sample of 39 AGN at $z\sim1.55$. While it may be more appropriate for high-redshift sources, the sample is biased against low-mass systems and therefore the calibration may not necessarily be better than the other calibrations which do sample the low-mass regime appropriate for \obj .}, \citet{Shen2011}, and \citet{Vestergaard2009} for \MgII\ and those from \citet{Mejia-Restrepo2016}, \citet{Vestergaard2006}, and \citet{McLure2004} for \hbeta . The calibration coefficients are:
\begin{equation}\label{eq:virialcoeff}
\begin{split}
    (a,\,b) &= (0.955, \,0.599), ~ {\rm M16};\,  {\rm Mg\,II} \\
    (a,\,b) &= (0.740, \,0.62), ~ {\rm S11};\, {\rm Mg\,II} \\
    (a,\,b) &= (0.860, \,0.50), ~ {\rm VO09};\, {\rm Mg\,II}  \\
    (a,\,b) &= (0.864, \,0.568), ~ {\rm M16};\, {\rm H}\beta \\
    (a,\,b) &= (0.910, \,0.50), ~ {\rm VP06};\, {\rm H}\beta \\
    (a,\,b) &= (0.672, \,0.61), ~ {\rm MD04};\, {\rm H}\beta .
\end{split}
\end{equation}

\autoref{tab:measurement} lists our results on the virial BH mass estimate. We estimate $M_{\bullet} {\sim}10^{6.43}$--$10^{6.72}M_{\odot}$ using broad \hbeta , or $M_{\bullet} {\sim}10^{7.14}$--$10^{7.36}M_{\odot}$ using broad \MgII\ based on the SDSS measurements. The range in the quoted BH mass estimate reflects the systematic uncertainty depending on the adopted calibrations. The total error in the BH mass estimate is dominated by systematic uncertainties in the virial mass estimates which are $\gtrsim$0.4 dex \citep[e.g.,][]{Shen2011}. This systematic uncertainty largely accounts for the fact that the empirically calibrated coefficients $a$ and $b$ may not necessarily apply to low-mass AGN at high redshift \citep[e.g.,][]{Grier2017}. \autoref{tab:measurement} also lists the BH mass estimates based on the OzDES measurements. We adopt the \hbeta -based value from SDSS as our fiducial estimate considering that \hbeta\ is better known and calibrated by reverberation mapping studies \citep[e.g.,][]{Shen2013} and is believed to be more reliable than \MgII\ as a virial mass estimator \citep[e.g.,][]{Guo2020a} and OzDES spectrum is incomplete for the \hbeta -\OIII\ region. 

We estimate the Eddington ratio $\lambda_{\rm Edd}\equiv \frac{L_{{\rm Bol}}}{L_{{\rm Edd}}}$ as $0.85\pm0.35$ for \obj\ from its hard X-ray luminosity $L_{\rm 2-10\,KeV}$ assuming a bolometric correction of $L_{{\rm Bol}}/L_{\rm 2-10\,KeV}=10$ \citep{Lusso2012L}. Considering the maximum $g-$band variability of 0.5 mag in Figure \ref{fig:lc}, \obj\ is consistent with the variability-Eddington ratio relation (see their Figure 11) in \citet{Rumbaugh2017}, which is produced with normal SDSS quasars of $M_{\rm BH} \approx 10^{9}M_{\odot}$.

\subsection{Host Galaxy Stellar Mass Estimation}\label{subsec:host_mass}

We estimate the host galaxy stellar mass by modeling its multi-wavelength spectral energy distribution (SED) using the software CIGALE\footnote{\url{https://cigale.lam.fr/about/}} \citep{Noll2009,Serra2011,Boquien2019}. CIGALE is designed to reduce computation time and the results are dependent on the parameter space explored by discrete models which can have degenerate physical parameter values. Mock catalogues are generated and analysed to check the reliability of estimated physical quantities.
%Mock catalogues are generated and analysed to mitigate the limitation of restricted parameter space. 

We assume an exponential ``delayed'' star formation history and vary the $e$-folding time and age of the stellar population model assuming solar metallicity and Chabrier initial mass function \citep[IMF;][]{chabrier03} to fit the stellar component. We adopt the single stellar population library from \citet{bc03} for the intrinsic stellar spectrum. We use templates from \citet{Inoue2011} based on CLOUDY 13.01 to model the nebular emission and amount of Lyman continuum photons absorbed by dust. We assume the dust attenuation curve of \citet{calzetti00} and a power law slope of $0$ to model dust attenuation. We model the dust emission using the empirical templates from \citet{Draine2007} with updates from \citet{Draine2014}. We use the templates from \citet{Fritz2006} to estimate the contribution from the AGN to the bolometric luminosity. The fractional contribution was allowed to vary from $0.1$ to $0.9$ along with the option for the object to be either type-1 or type-2 AGN.

Figure \ref{fig:sed} shows the SED data and our best-fit model. The best fit shown is for a type-1 AGNs with fractional contribution of 0.1 from the AGN to the bolometric luminosity\footnote{We caution that the SED photometries are measured at different times. This may introduce extra uncertainty to the estimation of the AGN component, considering the variability in \obj. In particular, the UV data points are sensitive to the AGN emission component.}. The resulting stellar mass estimate $M_{\ast}=10^{10.5\pm0.3}M_{\odot}$ can have around 20\% systematic uncertainty. More details about accuracy of estimating physical parameters related to stellar mass and fractional AGN contribution can be found in \citet{Boquien2019} and \citet{Ciesla2015}. 

To further quantify systematic uncertainties in our stellar mass estimate, we have double checked our result by fitting the SED using the software Prospector\footnote{\url{https://prospect.readthedocs.io/en/latest/index.html}} \citep{Leja2018}. Prospector is designed as a new framework for alleviating the model degeneracy and obtaining more accurate, unbiased parameters using the flexible stellar populations synthesis stellar populations code by \citet{Conroy2009}. SED fitting with both broad band photometries and spectroscopies are available in Prospector. Our best-fit stellar mass estimate from the Prospector analysis is $M_{\ast}=10^{10.8\pm0.5}M_{\odot}$, which is consistent with our CIGALE-based estimate within uncertainties.

\subsection{AGN Classification Using the Mass Excitation Diagnostics}\label{subsec:mex}

\begin{figure}
 \centering
 \includegraphics[width=0.5\textwidth]{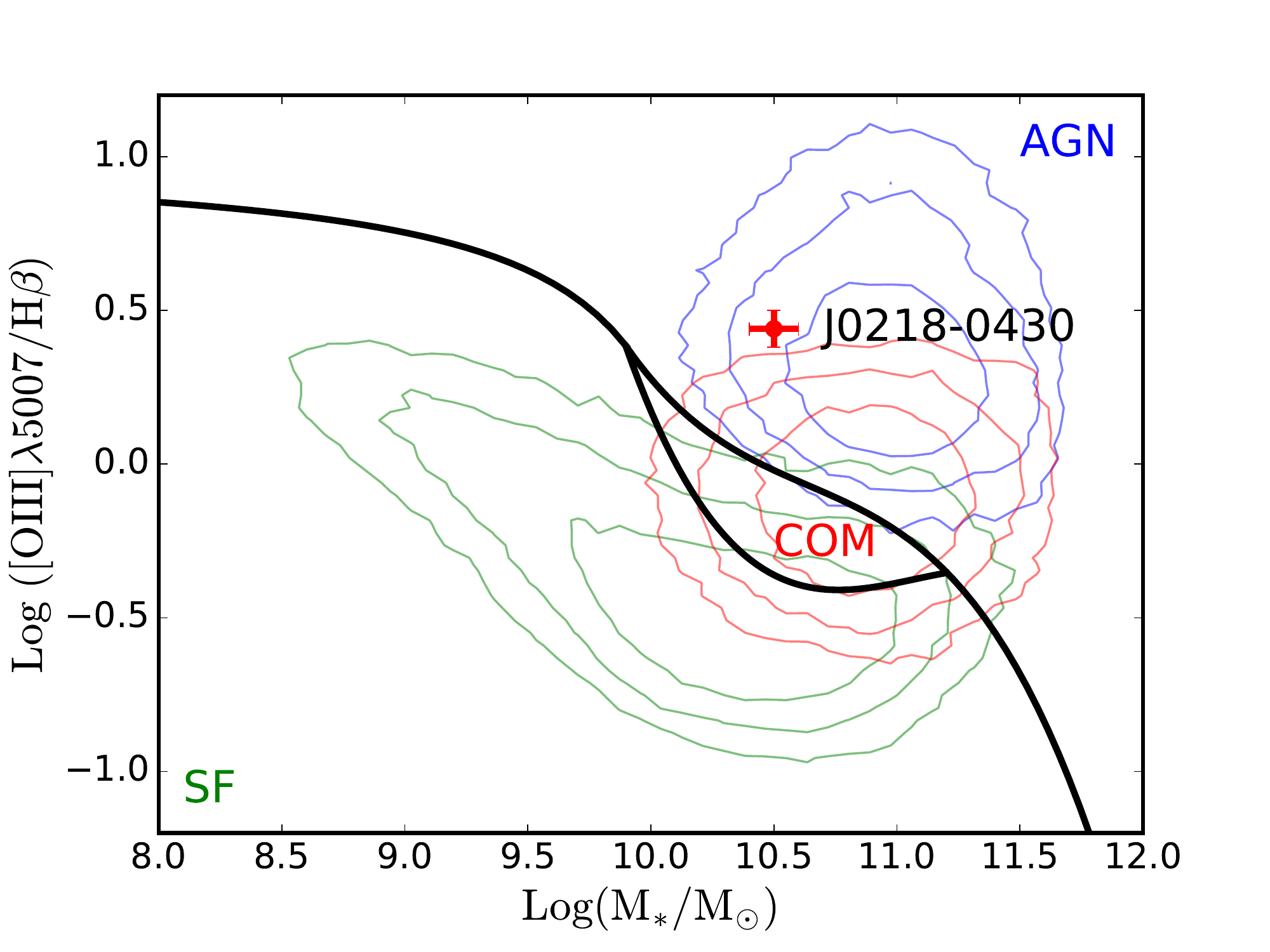}
    \caption{Mass excitation diagnostics for \obj . The black lines are boundaries defined by \citet{Juneau2011} to separate AGNs and star-forming galaxies. The green, red, and blue color contours represent number densities of pure star-forming galaxies, composites, and AGNs classified by the BPT diagram \citep{Kewley2001,Kauffmann2003}.
    }
    \label{fig:mex}
\end{figure}

Figure \ref{fig:mex} shows the mass excitation diagnostics diagram for \obj . This verifies that the gas excitation as inferred from the narrow emission-line ratio \OIII /\hbeta\ is dominated by the AGN rather than star formation. This is in line with the host galaxy being dominated by old stellar populations as suggested by the SED fitting. The mass excitation diagnostics provide further verification of the AGN classification in addition to direct evidence from the broad-line detection and the hard X-ray luminosity.

\section{Discussion}\label{sec:discuss}

\subsection{Comparison to Low-Mass AGNs in the Literature}\label{subsec:comparison}

\begin{figure*}
 \centering
 \includegraphics[width=1.0\textwidth]{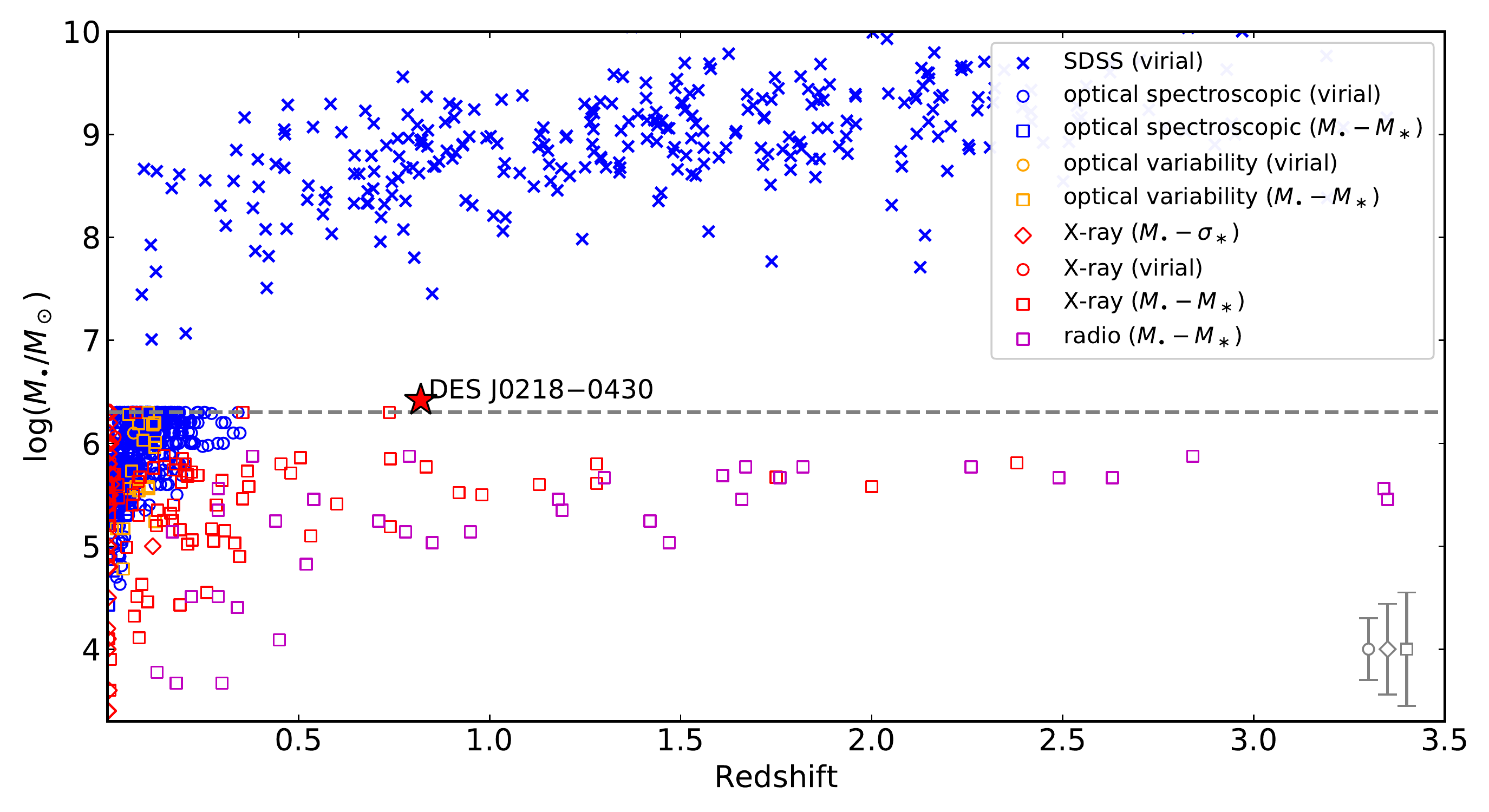}
    \caption{BH mass versus redshift for \obj\ in comparison to optical and X-ray-selected low-mass AGN candidates in the literature \citep{Filippenko2003,Barth2004,Greene2004,Greene2007a,Reines2011,Dong2012a,Ho2012,Secrest2012,Schramm2013,Reines2013,Maksym2014a,Baldassare2015,Lemons2015,Reines2015,Kawamuro2016,Pardo2016,Chang2017,She2017,Baldassare2018,Chilingarian2018,Ding2018,Liu2018,Mezcua2018} as well as the low-mass AGNs of the \citet{Mezcua2019} radio sample. This demonstrates \obj\ to be one of the lowest BH mass objects at similar redshift. The higher redshift X-ray selected sources are from the Chandra deep field. Additionally, \obj\ is the highest redshift object in its class identified from an optical survey. We consider objects with BHs mass estimates of $M_\bullet \leq 2\times10^{6}M_\odot$ and \obj. For comparison, the more massive sample of SDSS AGNs with BH masses from \citet{Shen2011} is also shown as blue crosses above the dashed line. The typical BH mass uncertainties are shown in grey at the lower right for $M_{\bullet}-M_{\ast}$ host scaling relation (0.3 dex), the virial method (0.44 dex), and the $M_{\bullet}-\sigma_{\ast}$ relation (0.55 dex). See \S4.1 (and references within) for details.
    }
    \label{fig:redshift_mass}
\end{figure*}

%We compiled a list of observed low-mass AGN candidates at varying redshift from references \citet{2017ApJ...842..131S,2003ApJ...588L..13F,2004ApJ...607...90B,2004ApJ...610..722G,2007ApJ...670...92G,2013ApJ...775..116R,2018ApJ...863....1C,2015ApJ...809L..14B,2015ApJ...813...82R,2018ApJS..235...40L,2018ApJ...868..152B,2016ApJ...831...37K,2012ApJ...761...73D,2018MNRAS.478.2576M,2017ApJS..233...19C,2018ApJ...868...88D,2011Natur.470...66R,2012ApJ...753...38S,2012ApJ...754...11H,2013ApJ...773..150S,2014MNRAS.444..866M,2015ApJ...805...12L,2016ApJ...831..203P}, shown in Fig.~\ref{fig:redshift_mass}. We do not include candidates with only Eddington-estimated luminosity BH mass lower limits. Many of these reference include candidates studied previously in earlier works. We removed duplicate entries, taking the virial estimations if available. Individual candidates may have differing BH mass estimates depending on the estimation method and techniques used. Therefore, the individual references should be consulted for details. When measured BH masses are not available, we use the $M_\bullet$--$M_\ast$ host scaling relation from \citet{2015ApJ...813...82R} to roughly estimate the BH mass.

Figure \ref{fig:redshift_mass} shows the BH mass versus redshift for \obj\ compared against a list of low-mass AGN candidates at different redshift compiled from the literature selected using various techniques. This demonstrates \obj\ as one of the lowest BH mass objects at similar redshift\footnote{Our fiducial BH mass is based on broad \hbeta\ which is believed to be more reliable than \MgII . In comparison, the AGN SDSS J021339.48$-$042456.4 at redshift $z=0.656$ has an estimated BH mass of $10^{7.83}M_{\odot}$ from \hbeta\ or $10^{6.43}M_{\odot}$ from \MgII\ \citep{Sanchez-Saez2018}.}. The comparison of \obj\ and known low-mass AGNs in the literature highlights the prospect of using optical variability in deep synoptic surveys to select low-mass AGNs toward higher redshift.

At similar redshifts to \obj, all low-mass AGN candidates in the literature are selected from X-ray deep-fields. We complied BH masses and redshifts for the samples noted in the figure caption. We removed duplicate entries during our literature search. We plot the virial BH mass measurements where possible. Individual candidates may have differing BH mass estimates depending on the estimation method and techniques used. Therefore, the individual references should be consulted for details. When measured BH masses are not available, we use the $M_\bullet$--$M_\ast$ host scaling relation from \citet{Reines2015} to estimate the BH mass. Although there are claims that these scaling relations may flatten-out below $M_\ast\sim10^{10}M_\odot$  \citep[e.g.,][]{Martin-Navarro2018} in addition to their large scatter, emphasizing the importance of obtaining broad-line BH mass measurements of low-mass AGNs.

\subsection{Comparison to Previous Optical and Near-IR Variability Searches of Low-Mass AGN}

\citet{Baldassare2018} used SDSS to select low-mass AGNs ($M_{*}\sim10^9{-}10^{10}M_{\odot}$) with a similar mass range as \obj\ but was limited to $z<0.15$. Our identification of a low-mass AGN at $z=0.823$ is enabled by the factor of 10 increase in single-epoch imaging sensitivity offered by DES-SN and detailed stellar mass estimation beyond the redshift limits of most stellar mass catalogs.

\citet{Martinez-Palomera2020} used DECam imaging to select galaxies with small amplitude ($g<0.1$ mag) variability characteristic of low-mass AGNs with no stellar mass cut. They confirm three AGNs with broad emission from SDSS spectroscopy in the range $M_{\bullet}\sim10^{6.0}{-}10^{6.5}M_{\odot}$. However, their sample is limited to $z<0.35$.

\citet{Sanchez-Saez2019} used a random-forest classifier trained on optical light curves (variable features and colors) using the QUEST-La Silla AGN variability survey with high purity. Their sample is dominated by quasars. These authors report the identification of eight low-luminosity AGNs which would not have been found with pure color selection or other traditional techniques. However, robust BH masses are not quoted in this work.

\citet{DeCicco2019} used the VST survey to select variable AGNs in the COSMOS field. This work also demonstrates variability selection is able to find AGNs with X-ray counterparts missed by color selection, but BH mass estimates are not reported for their sample.

\citet{Elmer2020} recently used NIR variability selection using \emph{K}-band imaging with the UKIDSS Ultra Deep Survey. These authors demonstrate the very valuable capability of NIR variability to identify AGNs in $M_{*}\sim10^9{-}10^{10}M_{\odot}$ hosts galaxies up to $z\sim3$, however BH mass estimates are not reported and virial BH masses are increasingly difficult to obtain for high-redshift low-mass AGNs.

\subsection{Implications for the BH-Host Scaling Relation at $z{\sim}1$}

\begin{figure}
 \centering
 \includegraphics[width=0.5\textwidth]{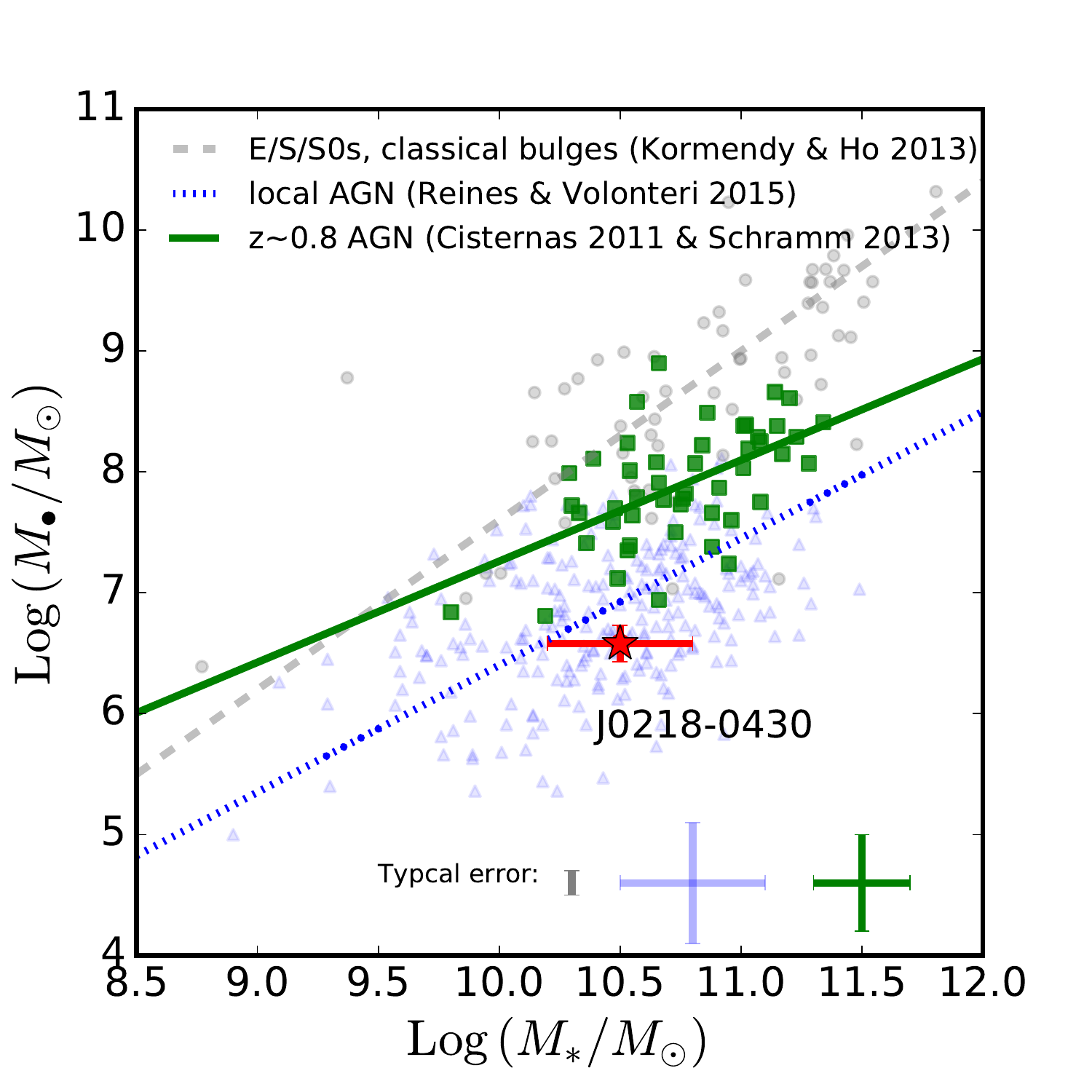}
    \caption{Black hole mass versus host-galaxy total stellar mass for \obj\ in comparison to X-ray selected intermediate-redshift AGN and local samples of AGN and inactive galaxies. The green solid line shows the best-fit relation of the sample of 48 X-ray selected AGN with a median $z\sim0.8$ from \citet{Cisternas2011} and \citet{Schramm2013} re-analyzed by \citet{Ding2020}. The blue dotted line represents the best-fit relation in local AGN from \citet{Reines2015} where the blue triangles show individual objects. The gray dashed line denotes the best-fit relation using the sample of ellipticals and spiral/S0 galaxies with classical bulges from \citet{KormendyHo2013} with the gray dots showing individual systems. 
   % The gray lines represent local scaling relations using the bulge stellar mass \citep{Haring2004,KormendyHo2013,McConnell2013}. 
    The error bars of \obj\ includes both statistical and systematic uncertainties. The error bars shown in the lower right corner denote typical uncertainties for the individual measurements in the comparison samples.
    }\label{fig:mass_mass}
\end{figure}

Figure \ref{fig:mass_mass} shows the virial BH mass versus host galaxy stellar mass for \obj . Shown for comparison is the X-ray selected AGN sample at median $z\sim0.8$ from \citet{Cisternas2011,Schramm2013} re-analyzed by \citet{Ding2020}. The virial BH masses were estimated based on single-epoch spectra using broad H$\beta$ and/or broad \MgII . The comparison sample includes 32 objects from \citet{Cisternas2011} and 16 objects from \citet{Schramm2013}. The total stellar masses of the \citet{Cisternas2011} sample were estimated by the empirical relation between $M_{*}/L$ and redshift and luminosity in the Hubble Space Telescope (HST) F814W band, which was established using a sample of 199 AGN host galaxies. The total stellar masses for the \citet{Schramm2013} sample were estimated from the galaxy absolute magnitude $M_{V}$ and rest-frame $(B-V)$ color measured from HST imaging for quasar-host decomposition using the $M/L$ calibration of \citet{Bell2003}. \obj\ extends the $M_{\bullet}$-$M_{\ast}$ relation at $z\sim1$ to smaller BH masses. \obj\ seems to have a BH mass $\sim3\sigma$ smaller than the median value we would expect from its total stellar mass. This may indicate that variability selection may identify AGNs with lower masses than X-ray selected AGN, although a larger sample is needed to draw a firm conclusion. Note that we also have assumed that low-mass AGNs usually reside in low-mass galaxies in our sample selection.

Also shown for context in Figure \ref{fig:mass_mass} are the best-fit scaling relations for local samples of inactive galaxies \citep[e.g.,][]{Haring2004,KormendyHo2013,McConnell2013} and low-redshift AGNs \citep{Reines2015}. While \obj\ appears to fall below the best-fit relation of low-redshift AGNs of \citet{Reines2015}, the apparent offset is insignificant accounting for systematic uncertainties in the virial BH mass estimate ($\sim0.44$ dex at 1 $\sigma$; \citealt{Shen2013}). While based on only one data point, our result on \obj\ suggests no significant redshift evolution in the $M_{\bullet}$--$M_{\ast}$ scaling relation from redshift $z{\sim}1$ to $z{\sim}0$ \citep[see also][]{Ding2020}, which is consistent with previous results based on the $M_{\bullet}$--$\sigma_{\ast}$ relation \citep[e.g.,][]{Shen2015,Sexton2019}. 

\section{Conclusion and Future Work}\label{sec:sum}

We have identified a low-mass AGN in the redshift $z=0.823$ galaxy \obj\ in DES-SN fields based on characterizing its long-term optical variability alone (Figures \ref{fig:lc}--\ref{fig:baseline}). We have not applied any color selection criterion to avoid bias induced by host galaxy starlight which dominates the optical to near-IR SED (Figure \ref{fig:sed}). We have confirmed the AGN nature by detecting broad \hbeta\ and broad \MgII\ in its archival optical spectrum (Figure \ref{fig:boss_spec}) from the SDSS-IV/eBOSS survey and by measuring its high X-ray 2--10 keV luminosity using archival XMM-Newton observations (\S \ref{subsec:sed_data}). We have estimated its virial BH mass as $M_{\bullet}\sim10^{6.43}$--$10^{6.72}M_{\odot}$ based on broad \hbeta\ from the SDSS (\S \ref{subsec:bh_mass}) and its host-galaxy stellar mass as $M_{\ast}=10^{10.5\pm0.3}M_{\odot}$ based on SED modeling (\S \ref{subsec:host_mass}). Comparing \obj\ to local samples of inactive galaxies and low-redshift AGN, we do not see any evidence for significant redshift evolution in the $M_{\bullet}$--$M_{\ast}$ relation from $z\sim1$ to $z\sim0$ (Figure \ref{fig:mass_mass}).  

\obj\ is one of the lowest BH mass objects at similar redshift (Figure \ref{fig:redshift_mass}). At similar redshifts to \obj, the literature IMBH candidates are all selected from X-ray deep-fields. Our work highlights the prospect of using optical variability to identify low-mass AGNs at higher redshift \citep[see also][for a recent study based on NIR variability]{Elmer2020}. 

In future work we will present a systematic variability search of all high-redshift low-mass AGN candidates in the DES-SN and deep fields. We will also systematically search for IMBHs using variability in low-redshift dwarf galaxies over the entire DES wide field based on low-cadence but long-term optical light curves. We will measure the black hole occupation functions and particularly at low masses to distinguish seed formation mechanisms. Finally, future observations with LSST will discover more small BHs at higher redshift as the more ``pristine'' fossil record to study BH seed formation.

\section*{Acknowledgements}

% especially thank the anonymous referee for his/her helpful comments and suggestions that have significantly improved the paper. 
We thank R. Kessler for help with getting the DES-SN light curve data, T. Davis and J. Hoormann for helpful correspondence, M. Mezcua, T. Diehl, S. Dodelson, T. Jeltema, and L. Whiteway for helpful comments, and the anonymous referee for a quick and careful report that improved the clarity of the paper. C.J.B. acknowledges support from the Illinois Graduate Survey Science Fellowship. Y.S. acknowledges support from the Alfred P. Sloan Foundation and NSF grant AST-1715579. 

Funding for DES Projects has been provided by the U.S. Department of Energy, the U.S. National Science Foundation, the Ministry of Science and Education of Spain, the Science and Technology Facilities Council of the United Kingdom, the Higher Education Funding Council for England, the National Center for Supercomputing Applications at the University of Illinois at Urbana-Champaign, the Kavli Institute of Cosmological Physics at the University of Chicago, the Center for Cosmology and Astro-Particle Physics at the Ohio State University, the Mitchell Institute for Fundamental Physics and Astronomy at Texas A\&M University, Financiadora de Estudos e Projetos, Funda{\c c}{\~a}o Carlos Chagas Filho de Amparo {\`a} Pesquisa do Estado do Rio de Janeiro, Conselho Nacional de Desenvolvimento Cient{\'i}fico e Tecnol{\'o}gico and the Minist{\'e}rio da Ci{\^e}ncia, Tecnologia e Inova{\c c}{\~a}o, the Deutsche Forschungsgemeinschaft and the Collaborating Institutions in the Dark Energy Survey. 

The Collaborating Institutions are Argonne National Laboratory, the University of California at Santa Cruz, the University of Cambridge, Centro de Investigaciones Energ{\'e}ticas, Medioambientales y Tecnol{\'o}gicas-Madrid, the University of Chicago, University College London, the DES-Brazil Consortium, the University of Edinburgh, the Eidgen{\"o}ssische Technische Hochschule (ETH) Z{\"u}rich, Fermi National Accelerator Laboratory, the University of Illinois at Urbana-Champaign, the Institut de Ci{\`e}ncies de l'Espai (IEEC/CSIC), the Institut de F{\'i}sica d'Altes Energies, Lawrence Berkeley National Laboratory, the Ludwig-Maximilians Universit{\"a}t M{\"u}nchen and the associated Excellence Cluster Universe, the University of Michigan, the National Optical Astronomy Observatory, the University of Nottingham, The Ohio State University, the University of Pennsylvania, the University of Portsmouth, SLAC National Accelerator Laboratory, Stanford University, the University of Sussex, Texas A\&M University, and the OzDES Membership Consortium.

Based in part on observations at Cerro Tololo Inter-American Observatory, National Optical Astronomy Observatory, which is operated by the Association of  Universities for Research in Astronomy (AURA) under a cooperative agreement with the National Science Foundation.

The DES data management system is supported by the National Science Foundation under Grant Numbers AST-1138766 and AST-1536171. The DES participants from Spanish institutions are partially supported by MINECO under grants AYA2015-71825, ESP2015-66861, FPA2015-68048, SEV-2016-0588, SEV-2016-0597, and MDM-2015-0509, some of which include ERDF funds from the European Union. IFAE is partially funded by the CERCA program of the Generalitat de Catalunya. Research leading to these results has received funding from the European Research Council under the European Union's Seventh Framework Program (FP7/2007-2013) including ERC grant agreements 240672, 291329, and 306478. We  acknowledge support from the Australian Research Council Centre of Excellence for All-sky Astrophysics (CAASTRO), through project number CE110001020, and the Brazilian Instituto Nacional de Ci\^enciae Tecnologia (INCT) e-Universe (CNPq grant 465376/2014-2).

This manuscript has been authored by Fermi Research Alliance, LLC under Contract No. DE-AC02-07CH11359 with the U.S. Department of Energy, Office of Science, Office of High Energy Physics. The United States Government retains and the publisher, by accepting the article for publication, acknowledges that the United States Government retains a non-exclusive, paid-up, irrevocable, world-wide license to publish or reproduce the published form of this manuscript, or allow others to do so, for United States Government purposes.

We are grateful for the extraordinary contributions of our CTIO colleagues and the DECam Construction, Commissioning and Science Verification teams in achieving the excellent instrument and telescope conditions that have made this work possible. The success of this project also relies critically on the expertise and dedication of the DES Data Management group.

Funding for the Sloan Digital Sky Survey IV has been provided by the Alfred P. Sloan Foundation, the U.S. Department of Energy Office of Science, and the Participating Institutions. SDSS-IV acknowledges support and resources from the Center for High-Performance Computing at the University of Utah. The SDSS web site is www.sdss.org.

SDSS-IV is managed by the Astrophysical Research Consortium for the Participating Institutions of the SDSS Collaboration including the Brazilian Participation Group, the Carnegie Institution for Science, Carnegie Mellon University, the Chilean Participation Group, the French Participation Group, Harvard-Smithsonian Center for Astrophysics, Instituto de Astrof\'isica de Canarias, The Johns Hopkins University, Kavli Institute for the Physics and Mathematics of the Universe (IPMU) / University of Tokyo, Lawrence Berkeley National Laboratory, Leibniz Institut f\"ur Astrophysik Potsdam (AIP),  Max-Planck-Institut f\"ur Astronomie (MPIA Heidelberg), Max-Planck-Institut f\"ur Astrophysik (MPA Garching), Max-Planck-Institut f\"ur Extraterrestrische Physik (MPE), National Astronomical Observatories of China, New Mexico State University, New York University, University of Notre Dame, Observat\'ario Nacional / MCTI, The Ohio State University, Pennsylvania State University, Shanghai Astronomical Observatory, United Kingdom Participation Group,Universidad Nacional Aut\'onoma de M\'exico, University of Arizona, University of Colorado Boulder, University of Oxford, University of Portsmouth, University of Utah, University of Virginia, University of Washington, University of Wisconsin, Vanderbilt University, and Yale University.

Facilities: DES, Sloan, OzDES

%%%%%%%%%%%%%%%%%%%%%%%%%%%%%%%%%%%%%%%%%%%%%%%%%%



\bibliographystyle{mnras}
\interlinepenalty=10000
\bibliography{ref.bib}

%%%%%%%%%%%%%%%%%%%%%%%%%%%%%%%%%%%%%%%%%%%%%%%%%%

\section*{affliations}  
$^{1}$ Department of Astronomy, University of Illinois at Urbana-Champaign, 1002 W. Green Street, Urbana, IL 61801, USA\\
$^{2}$ National Center for Supercomputing Applications, 1205 West Clark St., Urbana, IL 61801, USA\\
$^{3}$ Department of Physics, University of Illinois at Urbana-Champaign,  1110  West  Green  Street,  Urbana,  IL  61801,USA\\
$^{4}$ The Research School of Astronomy and Astrophysics, Australian National University, ACT 2601, Australia\\
$^{5}$ Departamento de F\'isica Matem\'atica, Instituto de F\'isica, Universidade de S\~ao Paulo, CP 66318, S\~ao Paulo, SP, 05314-970, Brazil\\
$^{6}$ Laborat\'orio Interinstitucional de e-Astronomia - LIneA, Rua Gal. Jos\'e Cristino 77, Rio de Janeiro, RJ - 20921-400, Brazil\\
$^{7}$ Fermi National Accelerator Laboratory, P. O. Box 500, Batavia, IL 60510, USA\\
$^{8}$ Instituto de Fisica Teorica UAM/CSIC, Universidad Autonoma de Madrid, 28049 Madrid, Spain\\
$^{9}$ CNRS, UMR 7095, Institut d'Astrophysique de Paris, F-75014, Paris, France\\
$^{10}$ Sorbonne Universit\'es, UPMC Univ Paris 06, UMR 7095, Institut d'Astrophysique de Paris, F-75014, Paris, France\\
$^{11}$ Department of Physics \& Astronomy, University College London, Gower Street, London, WC1E 6BT, UK\\
$^{12}$ Centro de Investigaciones Energ\'eticas, Medioambientales y Tecnol\'ogicas (CIEMAT), Madrid, Spain\\
$^{13}$ INAF, Astrophysical Observatory of Turin, I-10025 Pino Torinese, Italy\\
$^{14}$ INAF-Osservatorio Astronomico di Trieste, via G. B. Tiepolo 11, I-34143 Trieste, Italy\\
$^{15}$ Institute for Fundamental Physics of the Universe, Via Beirut 2, 34014 Trieste, Italy\\
$^{16}$ Observat\'orio Nacional, Rua Gal. Jos\'e Cristino 77, Rio de Janeiro, RJ - 20921-400, Brazil\\
$^{17}$ Department of Physics, IIT Hyderabad, Kandi, Telangana 502285, India\\
$^{18}$ Department of Astronomy/Steward Observatory, University of Arizona, 933 North Cherry Avenue, Tucson, AZ 85721-0065, USA\\
$^{19}$ Jet Propulsion Laboratory, California Institute of Technology, 4800 Oak Grove Dr., Pasadena, CA 91109, USA\\
$^{20}$ Santa Cruz Institute for Particle Physics, Santa Cruz, CA 95064, USA\\
$^{21}$ Institut d'Estudis Espacials de Catalunya (IEEC), 08034 Barcelona, Spain\\
$^{22}$ Institute of Space Sciences (ICE, CSIC),  Campus UAB, Carrer de Can Magrans, s/n,  08193 Barcelona, Spain\\
$^{23}$ Department of Astronomy, University of Michigan, Ann Arbor, MI 48109, USA\\
$^{24}$ Department of Physics, University of Michigan, Ann Arbor, MI 48109, USA\\
$^{25}$ Department of Physics, Stanford University, 382 Via Pueblo Mall, Stanford, CA 94305, USA\\
$^{26}$ Kavli Institute for Particle Astrophysics \& Cosmology, P. O. Box 2450, Stanford University, Stanford, CA 94305, USA\\
$^{27}$ SLAC National Accelerator Laboratory, Menlo Park, CA 94025, USA\\
$^{28}$ School of Mathematics and Physics, University of Queensland,  Brisbane, QLD 4072, Australia\\
\clearpage
$^{29}$ Center for Cosmology and Astro-Particle Physics, The Ohio State University, Columbus, OH 43210, USA\\
$^{30}$ Department of Physics, The Ohio State University, Columbus, OH 43210, USA\\
$^{31}$ Center for Astrophysics $\vert$ Harvard \& Smithsonian, 60 Garden Street, Cambridge, MA 02138, USA\\
$^{32}$ Australian Astronomical Optics, Macquarie University, North Ryde, NSW 2113, Australia\\
$^{33}$ Lowell Observatory, 1400 Mars Hill Rd, Flagstaff, AZ 86001, USA\\
$^{34}$ Instituci\'o Catalana de Recerca i Estudis Avan\c{c}ats, E-08010 Barcelona, Spain\\
$^{35}$ Institut de F\'{\i}sica d'Altes Energies (IFAE), The Barcelona Institute of Science and Technology, Campus UAB, 08193 Bellaterra (Barcelona) Spain\\
$^{36}$ Universit\'es Clermont Auvergne, CNRS/IN2P3, LPC, F-63000 Clermont-Ferrand, France\\
$^{37}$ Kavli Institute for Cosmological Physics, University of Chicago, Chicago, IL 60637, USA\\
$^{38}$ Institute of Astronomy, University of Cambridge, Madingley Road, Cambridge CB3 0HA, UK\\
$^{39}$ Department of Astrophysical Sciences, Princeton University, Peyton Hall, Princeton, NJ 08544, USA\\
$^{40}$ Department of Physics and Astronomy, Pevensey Building, University of Sussex, Brighton, BN1 9QH, UK\\
$^{41}$ School of Physics and Astronomy, University of Southampton,  Southampton, SO17 1BJ, UK\\
$^{42}$ Brandeis University, Physics Department, 415 South Street, Waltham MA 02453\\
$^{43}$ Computer Science and Mathematics Division, Oak Ridge National Laboratory, Oak Ridge, TN 37831\\
$^{44}$ Max Planck Institute for Extraterrestrial Physics, Giessenbachstrasse, 85748 Garching, Germany\\
$^{45}$ Universit\"ats-Sternwarte, Fakult\"at f\"ur Physik, Ludwig-Maximilians Universit\"at M\"unchen, Scheinerstr. 1, 81679 M\"unchen, Germany\
% Don't change these lines
\bsp	% typesetting comment
\label{lastpage}
\end{document}